\documentclass[]{spie}

\usepackage{amsmath,amsfonts,amssymb}
\usepackage{graphicx}
\usepackage[colorlinks=true, allcolors=blue]{hyperref}
\usepackage{booktabs}

\title{Development of an adaptive optics testbed at MPIA for the ELT/Planetary Camera and Spectrograph (PCS)}

\author[a]{Raphaël Pourcelot}

\author[a]{Thomas Bertram}
\author[a]{Peter Bizenberger}
\author[a]{Markus Feldt}
\author[b]{Markus Kasper}
\author[a]{Silvia Scheithauer}
\author[b]{Stefan Ströbele}
\author[a]{Horst Steuer}
\affil[a]{Max-Planck-Institut für Astronomie, Königstuhl 17, 69117 Heidelberg, Germany}
\affil[b]{European Southern Observatory (ESO), Karl-Schwarzschild-Str. 2,
85748 Garching, Germany}

\authorinfo{Further author information: (Send correspondence to R.P)\\E-mail: rapourcelot@mpia.de, Telephone: +49 622 152 8374}

\begin{document} 
\maketitle

\begin{abstract}
The Planetary Camera and Spectrograph (PCS) is a proposed second-generation instrument for the Extremely Large Telescope (ELT), dedicated to the direct imaging and characterization of exoplanets. To meet its demanding science requirements, PCS will incorporate an extreme adaptive optics (AO) system, building upon the heritage of existing ELT AO instruments such as ELT/METIS, as well as high-contrast AO systems at the ELT and the VLT, including SPHERE and its upcoming upgrade, SAXO+. PCS development requires extensive research and development to advance critical AO technologies. In this work, we present the Max Planck Institute for Astronomy (MPIA) plan for a modular testbed to validate key components and control strategies. This testbed will integrate two deformable mirrors, including a DM prototype developed by Bertin-ALPAO in collaboration with ESO, with an estimated delivery in 2029. The facility will enable testing of different Fourier filtering wavefront sensors, including novel mask designs, while exploring different control architectures, such as woofer-tweeter configurations with a single wavefront sensor for both deformable mirrors or fully independent AO stages. Additionally, the testbed will leverage MPIA’s expertise in real-time computer development to experiment with advanced control strategies, including predictive control and machine learning-enhanced AO techniques. This contribution presents the current status of PCS development at MPIA, highlighting the ongoing R\&D efforts to mature its AO system for high-contrast imaging with the ELT.
\end{abstract}

\keywords{Adaptive optics, deformable mirror, extreme AO, high-contrast imaging, ELT, exoplanets, wavefront sensing, real-time control}

\section{INTRODUCTION}
\label{sec:intro}

The Extremely Large Telescope (ELT) will see first light within a few years, representing a major step in ground-based astronomy. Its 39-meter primary mirror will provide unprecedented light-collecting power and spatial resolution, making it a unique tool for the direct imaging and spectroscopic characterization of exoplanets. The first generation of ELT instruments, METIS \cite{Brandl2021, Brandl2024, Feldt2024}, MICADO \cite{Davies2016}, and HARMONI \cite{Thatte2021}, will already incorporate sophisticated adaptive optics (AO) systems, and the lessons learned from their development and operation will be invaluable for the next generation.

The Planetary Camera and Spectrograph (PCS)\cite{Kasper2021, Kasper2026} is a proposed second-generation ELT instrument, conceived to push the frontiers of high-contrast imaging beyond what the first-generation instruments can achieve~\cite{chauvin2026}. Building on the legacy of VLT/SPHERE \cite{Beuzit2019} and its planned upgrade SAXO+ \cite{Boccaletti2026}, PCS targets the direct detection and characterization of reflected light from Neptune- and sub-Neptune-sized planets, as well as self-luminous giant planets. To accomplish this, according to previous studies \cite{Kasper2010}, PCS requires an extreme adaptive optics (ExAO) system capable of delivering Strehl ratios of 85\% or better in $H$-band, requiring both a high temporal frequency around 3\,kHz and high-spatial frequency control. 

In this context, the European Southern Observatory (ESO) is collaborating with Bertin-ALPAO, based in Grenoble, France, to push further deformable mirror (DM) technology and develop a prototype with more than 13k-actuators, yielding at least 128 actuators in the pupil diameter, and matching the actuation speed of 3\,kHz. Once manufactured, this DM will be tested at the Max-Planck-Institut for Astronomy (MPIA) in Heidelberg.

This paper describes the ongoing efforts at MPIA to develop a modular AO testbed to validate the Bertin-ALPAO 13k DM subsystem, as well as and control strategies in the context of PCS. Section~\ref{sec:pcs} summarises the PCS science case and instrument requirements. Section~\ref{sec:dm_req} presents the deformable mirror (DM) requirements and the prototyping activities underway with Bertin-ALPAO. Section~\ref{sec:testbed} describes the planned testbed architecture. Section~\ref{sec:rtc} discusses the real-time computing challenges. Section~\ref{sec:next} outlines the next steps, and Section~\ref{sec:conclusion} concludes.

\section{THE PLANETARY CAMERA AND SPECTROGRAPH}
\label{sec:pcs}

\subsection{Science Case}

PCS is designed to address one of the open questions in modern astrophysics: the direct detection and atmospheric characterisation of extrasolar planets in reflected starlight. The primary science targets are Neptune- and sub-Neptune-sized planets, whose reflected light signatures are far more challenging to detect than those of self-luminous giant planets~\cite{chauvin2026}. PCS will also be able to study protoplanetary and debris discs at high angular resolution, complementing the science achievable with the first-generation ELT instruments.

To enable these observations, PCS will operate in the visible and near-infrared ($I$ through $K$ band) and will incorporate both imaging and spectroscopic capabilities. The combination of the ELT's large collecting area and PCS's ExAO system is expected to provide access to contrast levels that are entirely out of reach for existing facilities.

\subsection{Contrast Requirements}

The science case imposes stringent contrast requirements on the combined ELT+PCS system. Table~\ref{tab:contrast} summarises the contrast goals in $I$-band ($\lambda = 806$~nm) as a function of angular separation and target brightness. These values drive nearly every subsystem requirement within the instrument.

\begin{table}[htb]
\caption{PCS contrast requirements in $I$-band (806 nm) for different limiting magnitudes and angular separations from the host star.}
\label{tab:contrast}
\centering
\begin{tabular}{|l|c|c|c|}
\hline
Limiting $I$ mag & 30 mas ($\sim 8\,\lambda/D$) & 100 mas & 300 mas \\
\hline
5 (goal 6) & $2\times10^{-9}$ (goal $10^{-9}$) & $10^{-9}$ & $5\times10^{-10}$ (goal $2\times10^{-10}$) \\
7 (goal 8) & $2\times10^{-8}$ & $10^{-9}$ & $10^{-9}$ \\
9 (goal 10) & $10^{-6}$ & $10^{-6}$ & $10^{-6}$ \\
\hline
\end{tabular}
\end{table}

Achieving $2\times10^{-9}$ contrast at only $8\,\lambda/D$ in $I$-band represents an improvement of approximately two orders of magnitude over what SPHERE can currently deliver, and underscores the need for a dedicated ExAO development programme.

\subsection{AO Architecture Overview}

Current PCS studies point toward the use of a two-stage AO correction. The ELT's adaptive secondary mirror M4, which is common to all ELT instruments, will provide a first stage of correction. A dedicated high-order deformable mirror (DM) within PCS will then provide the second, instrument-specific ExAO correction. Overall, the AO loop must correct at frequencies of approximately 3~kHz with a total end-to-end latency of less than 500~$\mu$s (corresponding to roughly 1.5 frames), and must deliver $\geq85\%$ Strehl ratio in $H$-band.

\section{DEFORMABLE MIRROR REQUIREMENTS AND PROTOTYPING}

\subsection{Deformable mirror requirements}
\label{sec:dm_req}

Based on previous EPICS analysis,\cite{Kasper2010} the contrast and Strehl requirements translate directly into demanding specifications for the high-order DM. The requirement for more than 128 actuators across the pupil with a high Strehl ratio results in a number of key parameters. Mechanically, 1.5~mm actuator pitch, the $13\,000 $  actuators will fit into a pupil of approximately 200~mm in diameter. For the wavefront requirements it must achieve a static flatness better than 75~nm RMS wavefront error over the full aperture, together with a mechanical stroke of at least  $ 3\,\mu$m and a resolution of at least 0.2~nm per step. In terms of dynamic behavior, the mirror is required to move from rest to within  $ \pm 20\% $  of a 30~nm wavefront error target in less than  $ 300\,\mu $s, while operating at a frequency exceeding 3~kHz. Finally, the associated real-time computer latency budget is restricted to approximately $ 180\,\mu$s in order to remain within the total latency envelope.

These specifications, and in particular the combination of high actuator count, large stroke, fast rise time, and sub-nanometer resolution, represent a significant technological challenge.

\subsection{Deformable mirror prototype development}

A DM prototype is being developed by Bertin-ALPAO (Grenoble, France) in collaboration with ESO and MPIA. Delivery of the prototype is anticipated in 2029. The prototype will serve as a risk-reduction activity ahead of the PCS final design, and will be used to characterise and validate the mirror's performance in a controlled laboratory environment at MPIA. The characterisation campaign at MPIA will cover, among other properties, the rise speed and the detailed profile of the temporal response, as well as the static and dynamic precision and accuracy of the mirror. The campaign will also assess the overshoot and undershoot behaviour, together with any rippling and mechanical resonances that appear during representative AO control sequences. Finally, the closed-loop AO behaviour of the mirror will be evaluated within the full testbed, to mimic a second-stage AO loop downstream of a first stage correction made by the ELT deformable mirror M4. 

A key diagnostic for the precision measurements is the photon budget per actuator per frame. Achieving 1~nm precision per actuator per frame requires of order $15\,000$ detected photons per exposure, which also places requirements on the light source brightness and wavefront sensor throughput. 

\section{THE MPIA TESTBED FOR PCS AO}
\label{sec:testbed}

\subsection{Design goals}

he MPIA testbed will be designed with modularity in mind, allowing it to be reconfigured as the PCS design matures and as different subsystems become available for testing, such as an easy swap of the wavefront sensors between different Fourier-filtering ones, or the replacement of the Bertin-ALPAO DM by a flat reference for calibration. The primary objectives of the testbed are to characterize and qualify the Bertin-ALPAO DM prototype, and to explore AO loop strategies with this DM in closed-loop experiments. The testbed further aims to demonstrate high-contrast performance at the $2\times10^{-5}$ level at $5\lambda/D$, and to evaluate and compare different wavefront sensors, such as the pyramid,\cite{Ragazzoni1996} Zernike,\cite{Bloemhof2003} Bi-O Edge,\cite{Verinaud2024} or integrated Mach-Zehnder\cite{Graf2025} concepts. Finally, the testbed is intended to support the development and testing of real-time control (RTC) software and hardware capable of operating at the required speed and scale.

\subsection{Optical Layout}

\begin{figure}[htb]
    \centering
    \includegraphics[width=\columnwidth]{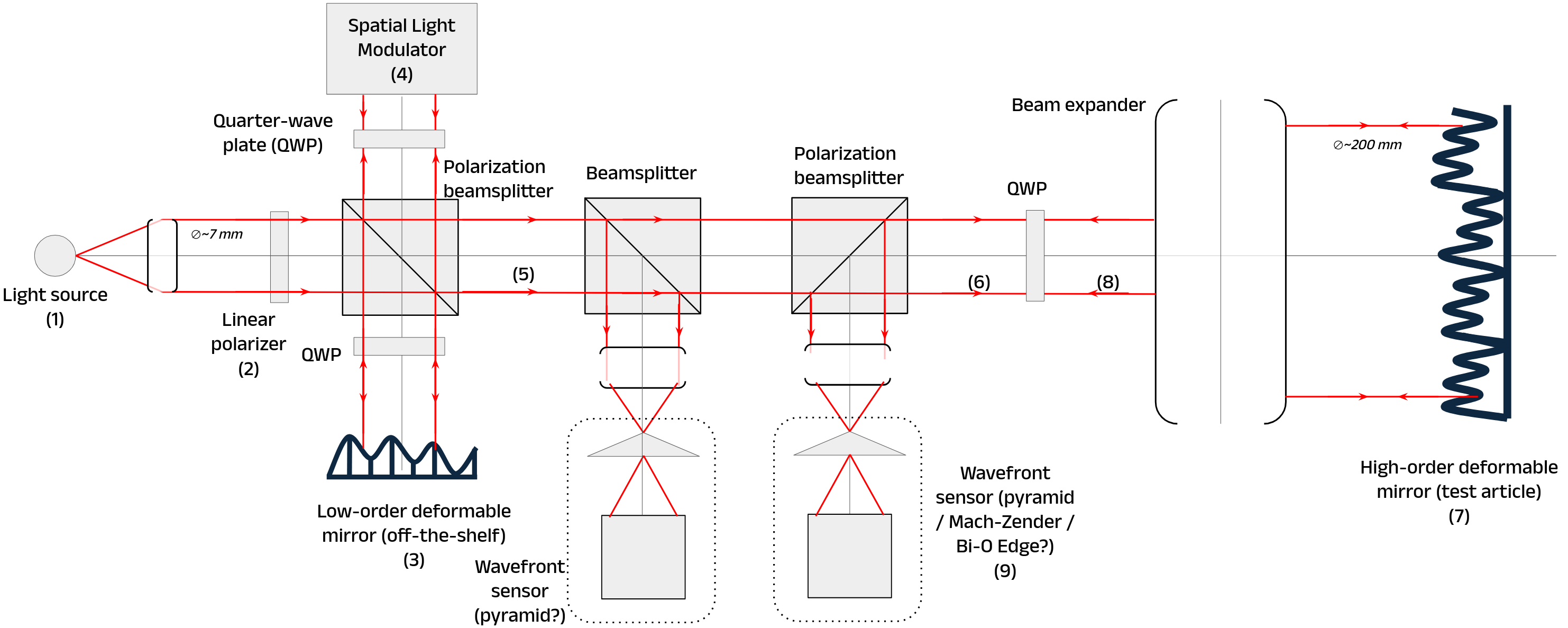}
    \caption{Early draft of a possible layout for the MPIA to test the Bertin-Alpao 13k DM. Key components are numbered and described in the text. This version makes use of polarization components to simplify the number of optical elements.}
    \label{fig:testbed}
\end{figure}

We present here an early draft of the testbed layout, with a schematic overview shown in Fig.~\ref{fig:testbed}. This version designed around a set of polarization beamsplitters and quarter-wave plates that act as optical isolators, enabling multiple passes through the same optics, keeping the number of optics to a minimum. The order in which the beam is propagating is indicated with numbers, that are also detailed in this section. This layout consists in two separated sections. One with a smaller beam of diameter smaller than 10 mm, that matches off-the-shelf low-order deformable mirrors. The other one is after the beam expander, to match the diameter of the high-order DM prototype. Note that not all challenges have been solved yet. 

The main optical components are as follows. The light source~(1) is a point-like source, which could be a super-luminescent diode chosen for its brightness and narrow bandpass. A linear polarizer~(2) is used to define the input polarization state for the polarimetric beam routing described below, and that, alongside with the quarter-wave plates and the polarization beamsplitters, for optical isolators. These routes the beam through the DMs while suppressing back-reflections and facilitating beam extraction without additional beamsplitters. A low-order, off-the-shelf deformable mirror~(3) acts as the woofer, providing pre-compensation of large-amplitude, low-spatial-frequency aberrations, and would also enable tests to validate a woofer-tweeter control topology. A spatial light modulator~(4) is used for disturbance injection with a high spatial component, enabling the simulation of static and dynamic wavefront aberrations representative of atmospheric turbulence and telescope aberrations; the exact model has not yet been chosen and will likely drive the speed of the injected turbulence. A dedicated pre-compensator wavefront sensor monitors the output of the low-order DM to enable independent control of the first stage. The high-order DM prototype from Bertin-ALPAO~(7) occupies the following position and is the primary component under test. Finally, the wavefront sensor position~(8) is designed to accommodate multiple wavefront sensor concepts, including pyramid wavefront sensors, Mach-Zehnder interferometers, or a Bi-O Edge sensor. 

\subsection{Wavefront Sensor Trade-offs}

A key scientific objective of the testbed is to compare different wavefront sensor architectures relevant to ExAO. The pyramid wavefront sensor (PWFS) is one of the leading candidate for PCS, given its demonstrated sensitivity advantage over the Shack-Hartmann sensor in the photon-noise-limited regime. However, alternative Fourier-filtering wavefront sensors, including Mach-Zehnder-based designs and the Bi-O Edge sensor, will also be evaluated. The testbed is designed to allow the wavefront sensor module to be exchanged with minimal disruption to the rest of the system.

\subsection{Control Architecture}

The testbed will also enable the investigation of different control architectures with the Bertin-ALPAO high-order DM in a two-DM system. The first one is with two independent AO stages, where each DM is driven by its own dedicated wavefront sensor, operating independently. This architecture is conceptually simpler to control but requires additional hardware. The second one is in a woofer-tweeter way with a single wavefront sensor, used to drive both DMs simultaneously. The control signals are decomposed in modal space, with low spatial frequencies directed to the woofer DM and high spatial frequencies to the tweeter DM. This architecture minimizes hardware complexity but requires careful modal separation to avoid cross-talk and instability.
Both architectures will be evaluated in terms of closed-loop Strehl ratio, contrast, and robustness to misregistration and calibration errors.

\subsection{Photon budget}
For a pyramid wavefront sensor operating in the photon-noise-limited regime and without readout noise, the measurement error $\sigma$ for a mode $\phi$ is given by $\sigma_\phi = \lambda / (2\pi\sqrt{N})$ \cite{Chambouleyron2023}, where $\lambda$ is the sensing wavelength and $N$ is the number of photons per frame for this mode. Inverting this relation yields the required photon count $N = (\lambda / 2\pi\sigma_\phi)^2$, which for a target accuracy of $\sigma_\phi < 1$ nm at $\lambda = 800$ nm gives $N \approx 16{,}200$ photons per pixel per frame per mode. Considering 13000 modes for the 13k DM, this translates into an optical power of $\sim 1 \mu$W at 3 kHz framerate, leaving some margin for more complex systems. 

\section{REAL-TIME COMPUTING CHALLENGES}
\label{sec:rtc}

Operating a system with more than $13\,000$ actuators at 3~kHz imposes extreme demands on the real-time computer. For a back-of-the-envelope calculation, the wavefront sensor, whether a pyramid or an alternative design, is assumed to have two pixels per actuator per sub-aperture across a $128\times128$ lenslet grid with four pupil images, giving a total of $2\times128\times128\times4 = 131\,072$ pixels per frame. In terms of pixel reception throughput, at 3~kHz with 16-bit pixels the camera link must sustain approximately 786~MB/s. Regarding the command matrix, in the slopes-based control approach the reconstructor matrix relating WFS slopes to DM actuator commands has dimensions of approximately $13\,000\times52\,000$, corresponding to a matrix of about 1.4~GB in single precision. Performing this matrix-vector multiplication (MVM) at 3~kHz requires a sustained memory and arithmetic throughput of approximately 4~TB/s.

These figures will require special attention and dedicated hardware such as GPU-based computation. MPIA has been developing in-house RTC expertise, in particular for the ELT/METIS instrument, and this infrastructure will be directly leveraged for PCS.

\section{NEXT STEPS}
\label{sec:next}

The immediate near-term priorities for the MPIA testbed programme are the following. The first priority consists of extensive design studies to define a modular optical design that enables DM testing while allowing later evolutions of the testbed. The second is the procurement of off-the-shelf hardware, namely the acquisition of the low-order DM, the pre-compensator WFS, the SLM, and supporting opto-mechanical components required to commission the first iteration of the testbed. The third priority is testbed commissioning, comprising the alignment and characterization of the basic AO loop with the low-order DM and a pyramid WFS. The fourth is coronagraph integration, involving the addition of a simple coronagraph, such as a Lyot-type design, to the science arm, in order to enable contrast measurements at the $2\times10^{-5}$ level at $5\lambda/D$ as an initial validation milestone; this intermediate contrast goal is consistent with the requirements of the testbed development phase and will serve as a stepping stone toward the science requirements. Then will come the DM prototype integration, which will follow delivery of the Bertin-ALPAO prototype and consist of its integration into the testbed and the execution of the full characterization campaign described in Section~\ref{sec:dm_req}. Finally, the sixth priority is RTC development, carried out in parallel and targeting the throughput figures quoted in Section~\ref{sec:rtc}.

\section{CONCLUSION}
\label{sec:conclusion}

PCS represents the next frontier in direct exoplanet imaging from the ground. Its science requirements, with post-processed contrast ratios of $2\times10^{-9}$ at $8\,\lambda/D$ in $I$-band demand an ExAO system that goes beyond the current state of the art. The required DM, with $>13\,000$ actuators, sub-nanometer resolution, and a rise time of $<300\,\mu$s, does not yet exist as a commercial product, underscoring the need for a dedicated R\&D program.

MPIA is developing a modular AO testbed specifically designed to support this development. The testbed will integrate two deformable mirrors, including the Bertin-ALPAO DM prototype, in a flexible optical layout that accommodates multiple wavefront sensor and control architecture trade-offs. MPIA's existing RTC expertise will be extended to meet the extreme computational demands of the 13\,000-actuator, 3~kHz control loop.

This facility will play a central role in preparing the PCS ExAO system ahead of the instrument's final design phase, and will serve as a platform for testing advanced control strategies and architectures that are essential steps for PCS to deliver the science it is aiming for.
\acknowledgments
The authors acknowledge the use of Claude Opus for some part of text redaction, language edition, spelling and grammar check. The content and calculations did not involve AI.

\bibliography{report} 

@INPROCEEDINGS{Kasper2026,
       author = {{Kasper}, Markus},
        title = "{PCS - the ELT exoPlanet imaging Camera and Spectrograph}",
    booktitle = {Planetary formation and Exoplanets in the ELT era (Exo-ELT)},
         year = 2026,
        month = apr,
          eid = {33},
        pages = {33},
          doi = {10.5281/zenodo.19686421},
       adsurl = {https://ui.adsabs.harvard.edu/abs/2026exoe.confE..33K}
}

@ARTICLE{Kasper2021,
       author = {{Kasper}, M. and {Cerpa Urra}, N. and {Pathak}, P. and {Bonse}, M. and {Nousiainen}, J. and {Engler}, B. and {Heritier}, C.~T. and {Kammerer}, J. and {Leveratto}, S. and {Rajani}, C. and et al.},
        title = "{PCS {\textemdash} A Roadmap for Exoearth Imaging with the ELT}",
      journal = {The Messenger},
         year = 2021,
        month = mar,
       volume = {182},
        pages = {38-43},
          doi = {10.18727/0722-6691/5221},
archivePrefix = {arXiv},
       eprint = {2103.11196},
 primaryClass = {astro-ph.IM},
       adsurl = {https://ui.adsabs.harvard.edu/abs/2021Msngr.182...38K}
}

@unpublished{Chauvin2026,
    author = {{Chauvin}, G. et al.},
    title = {PCS Roadmap study - Science cases},
    year = 2026,

}

@ARTICLE{Beuzit2019,
       author = {{Beuzit}, J.-L. and {Vigan}, A. and {Mouillet}, D. and {Dohlen}, K. and {Gratton}, R. and {Boccaletti}, A. and {Sauvage}, J.-F. and {Schmid}, H.~M. and {Langlois}, M. and {Petit}, C. and et al.},
        title = "{SPHERE: the exoplanet imager for the Very Large Telescope}",
      journal = {Astronomy and Astrophysics},
         year = 2019,
        month = nov,
       volume = {631},
          eid = {A155},
        pages = {A155},
          doi = {10.1051/0004-6361/201935251},
archivePrefix = {arXiv},
       eprint = {1902.04080},
 primaryClass = {astro-ph.IM},
       adsurl = {https://ui.adsabs.harvard.edu/abs/2019A&A...631A.155B}
}

@INPROCEEDINGS{Boccaletti2026,
       author = {{Boccaletti}, Anthony},
        title = "{Upgrading SPHERE with the second stage AO system SAXO+: a technology demonstrator for PCS}",
    booktitle = {Planetary formation and Exoplanets in the ELT era (Exo-ELT)},
         year = 2026,
        month = apr,
          eid = {34},
        pages = {34},
          doi = {10.5281/zenodo.19686471},
       adsurl = {https://ui.adsabs.harvard.edu/abs/2026exoe.confE..34B}
}

@inproceedings{Brandl2024,
author = {Bernhard R. Brandl and Felix Bettonvil and Roy van Boekel and Adrian Glauser and Sascha P. Quanz and Olivier Absil and Markus Feldt and Paulo J. V. Garcia and Alistair Glasse and Manuel Guedel and Lucas Labadie and Michael Meyer and Eric Pantin and Shiang-Yu Wang and Hans van Winckel},
title = {{Final design and status of the Mid-IR ELT Imager and Spectrograph, METIS}},
volume = {13096},
booktitle = {Ground-based and Airborne Instrumentation for Astronomy X},
editor = {Julia J. Bryant and Kentaro Motohara and Jo{\"e}l R. D. Vernet},
organization = {International Society for Optics and Photonics},
publisher = {SPIE},
pages = {1309612},
year = {2024},
doi = {10.1117/12.3018975},
URL = {https://doi.org/10.1117/12.3018975}
}

@ARTICLE{Brandl2021,
       author = {{Brandl}, B. and {Bettonvil}, F. and {van Boekel}, R. and {Glauser}, A. and {Quanz}, S. and {Absil}, O. and {Amorim}, A. and {Feldt}, M. and {Glasse}, A. and {G{\"u}del}, M. and et al.},
        title = "{METIS: The Mid-infrared ELT Imager and Spectrograph}",
      journal = {The Messenger},
         year = 2021,
        month = mar,
       volume = {182},
        pages = {22-26},
          doi = {10.18727/0722-6691/5218},
archivePrefix = {arXiv},
       eprint = {2103.11208},
 primaryClass = {astro-ph.IM},
       adsurl = {https://ui.adsabs.harvard.edu/abs/2021Msngr.182...22B}
}

@ARTICLE{Feldt2024,
       author = {{Feldt}, Markus and {Bertram}, Thomas and {Correia}, Carlos and {Absil}, Olivier and {C{\'a}rdenas V{\'a}zquez}, M. Concepci{\'o}n and {Coppejans}, Hugo and {Kulas}, Martin and {Obereder}, Andreas and {Orban de Xivry}, Gilles and {Scheithauer}, Silvia and et al.},
        title = "{High strehl and high contrast for the ELT instrument METIS: Final design, implementation, and predicted performance of the single-conjugate adaptive optics system}",
      journal = {Experimental Astronomy},
         year = 2024,
        month = dec,
       volume = {58},
       number = {3},
          eid = {20},
        pages = {20},
          doi = {10.1007/s10686-024-09968-2},
archivePrefix = {arXiv},
       eprint = {2411.17341},
 primaryClass = {astro-ph.IM},
       adsurl = {https://ui.adsabs.harvard.edu/abs/2024ExA....58...20F}
}

@article{Thatte2021,
  doi = {10.18727/0722-6691/5215},
  
  url = {http://doi.eso.org/10.18727/0722-6691/5215},
  
  author = {Thatte, Niranjan and Tecza, Matthias and Schnetler, Hermine and Neichel, Benoit and Melotte, Dave and Fusco, Thierry and Ferraro-Wood, Vanessa and Clarke, Fraser and Bryson, Ian and O’Brien, Kieran and Mateo, Mario and Garcia Lorenzo, Begoña and Evans, Chris and Bouché, Nicolas and Arribas, Santiago and Consortium, The HARMONI},
  
  title = {HARMONI: the ELT’s First-Light Near-infrared and Visible Integral Field Spectrograph},
  
  journal = {Published in The Messenger vol. 182},
  
  volume = {pp. 7-12},
  
  pages = {March 2021.},
  
  publisher = {European Southern Observatory (ESO)},
  
  year = {2021},
  
  copyright = {Copyright European Southern Observatory}
}

@INPROCEEDINGS{Davies2016,
       author = {{Davies}, R. and {Schubert}, J. and {Hartl}, M. and {Alves}, J. and {Cl{\'e}net}, Y. and {Lang-Bardl}, F. and {Nicklas}, H. and {Pott}, J.-U. and {Ragazzoni}, R. and {Tolstoy}, E. and et al.},
        title = "{MICADO: first light imager for the E-ELT}",
    booktitle = {Ground-based and Airborne Instrumentation for Astronomy VI},
         year = 2016,
       editor = {{Evans}, Christopher J. and {Simard}, Luc and {Takami}, Hideki},
       series = {Society of Photo-Optical Instrumentation Engineers (SPIE) Conference Series},
       volume = {9908},
        month = aug,
          eid = {99081Z},
        pages = {99081Z},
          doi = {10.1117/12.2233047},
archivePrefix = {arXiv},
       eprint = {1607.01954},
 primaryClass = {astro-ph.IM},
       adsurl = {https://ui.adsabs.harvard.edu/abs/2016SPIE.9908E..1ZD}
}

@INPROCEEDINGS{Kasper2010,
       author = {{Kasper}, Markus and {Beuzit}, Jean-Luc and {Verinaud}, Christophe and {Gratton}, Raffaele G. and {Kerber}, Florian and {Yaitskova}, Natalia and {Boccaletti}, Anthony and {Thatte}, Niranjan and {Schmid}, Hans Martin and {Keller}, Christoph and et al.},
        title = "{EPICS: direct imaging of exoplanets with the E-ELT}",
    booktitle = {Ground-based and Airborne Instrumentation for Astronomy III},
         year = 2010,
       editor = {{McLean}, Ian S. and {Ramsay}, Suzanne K. and {Takami}, Hideki},
       series = {Society of Photo-Optical Instrumentation Engineers (SPIE) Conference Series},
       volume = {7735},
        month = jul,
          eid = {77352E},
        pages = {77352E},
          doi = {10.1117/12.856850},
       adsurl = {https://ui.adsabs.harvard.edu/abs/2010SPIE.7735E..2EK}
}

@article{Verinaud2024,
	author = {{V\'erinaud, C.} and {H\'eritier, C. T.} and {Kasper, M.} and {Tallon, M.}},
	title = {The Bi-O edge wavefront sensor - How Foucault-knife-edge variants can boost extreme adaptive optics},
	DOI= "10.1051/0004-6361/202346660",
	url= "https://doi.org/10.1051/0004-6361/202346660",
	journal = {A\&A},
	year = 2024,
	volume = 682,
	pages = "A27",
}

@ARTICLE{Chambouleyron2023,
       author = {{Chambouleyron}, V. and {Fauvarque}, O. and {Plantet}, C. and {Sauvage}, J.-F. and {Levraud}, N. and {Ciss{\'e}}, M. and {Neichel}, B. and {Fusco}, T.},
        title = "{Modeling noise propagation in Fourier-filtering wavefront sensing, fundamental limits, and quantitative comparison}",
      journal = {Astronomy and Astrophysics},
         year = 2023,
        month = feb,
       volume = {670},
          eid = {A153},
        pages = {A153},
          doi = {10.1051/0004-6361/202245351},
archivePrefix = {arXiv},
       eprint = {2212.13577},
 primaryClass = {astro-ph.IM},
       adsurl = {https://ui.adsabs.harvard.edu/abs/2023A&A...670A.153C}
}

@article{Graf2025,
	author = {{Graf, C.} and {Langlois, M.} and {Thi\'ebaut, \'E.} and {Tallon, M.}},
	title = {Calibration and performances of the integrated Mach-Zehnder wavefront sensor for extreme adaptive optics},
	DOI= "10.1051/0004-6361/202553987",
	url= "https://doi.org/10.1051/0004-6361/202553987",
	journal = {A\&A},
	year = 2025,
	volume = 701,
	pages = "A63",
}

@ARTICLE{Ragazzoni1996,
       author = {{Ragazzoni}, Roberto},
        title = "{Pupil plane wavefront sensing with an oscillating prism}",
      journal = {Journal of Modern Optics},
         year = 1996,
        month = feb,
       volume = {43},
       number = {2},
        pages = {289-293},
          doi = {10.1080/09500349608232742},
       adsurl = {https://ui.adsabs.harvard.edu/abs/1996JMOp...43..289R}
}

@INPROCEEDINGS{Bloemhof2003,
       author = {{Bloemhof}, Eric E. and {Wallace}, J. Kent},
        title = "{Phase contrast techniques for wavefront sensing and calibration in adaptive optics}",
    booktitle = {Astronomical Adaptive Optics Systems and Applications},
         year = 2003,
       editor = {{Tyson}, Robert K. and {Lloyd-Hart}, Michael},
       series = {Society of Photo-Optical Instrumentation Engineers (SPIE) Conference Series},
       volume = {5169},
        month = dec,
        pages = {309-320},
          doi = {10.1117/12.507245},
       adsurl = {https://ui.adsabs.harvard.edu/abs/2003SPIE.5169..309B}
}
\bibliographystyle{spiebib}

\end{document}